\documentclass[aps,prb,twocolumn,longbibliography]{revtex4}

\usepackage{epsfig}
\usepackage{soul}

\usepackage{bm}
\usepackage{graphicx}
\usepackage{amsmath}
\usepackage{amssymb}
\usepackage[utf8]{inputenc}
\usepackage[T1]{fontenc}
\usepackage{color}
\usepackage{upgreek}
\usepackage{subfigure}
\usepackage{placeins}
\usepackage{lipsum}
\usepackage{epsfig}
\usepackage[utf8]{inputenc}
\usepackage{wrapfig}

\renewcommand{\Im}{\mathop{\rm Im}}

\newcommand{\effectname}{{CDOP}}

\begin{document}
\title{ The drag of photons by electric current in quantum wells\\
}

\author{G.V. Budkin$^1$, I.S. Makhov$^{2}$, D.A. Firsov$^{2}$
}

\affiliation{$^1$Ioffe Institute, 194021	St. Petersburg, Russia}

\affiliation{$^2$ Peter the Great St. Petersburg Polytechnic University, 195251, St. Petersburg, Russia}

\begin{abstract}
The flow of electric current in quantum well breaks the space inversion symmetry, which leads to
 the dependence of the radiation transmission on the relative orientation of current and photon wave vector, this phenomenon can be named current drag of photons.
We have developed a microscopic theory of such an effect for intersubband transitions in quantum wells taking into account both depolarization and exchange-correlation effects.
It is shown that the effect of the current drag of photons originates from
the asymmetry of intersubband optical transitions due to the redistribution of electrons in momentum space. We show that the presence of dc electric current leads to the shift of intersubband resonance position and affects both transmission coefficient and absorbance in quantum wells.
\end{abstract}

\maketitle

\section{Introduction}

A wide variety of optical and transport phenomena can be observed in semiconductor structures.
Often the determining factor in whether a particular phenomenon can be observed in a specific structure is the symmetry of this structure.
The presence of heterointerfaces in nanostructures by itself leads to a reduction of spatial symmetry
in comparison with bulk materials~\cite{Ivchenko1997}, and allows the emergence of new effects absent in bulk semiconductors.
The symmetry of the structure can be further controlled 
and lowered in various ways, e.g. deformation, a gradient of temperature, magnetic field, or dc electric current.
The latter can lead to a number of electro-optical effects, such as current-induced optical activity~\cite{vorobev1979,Tate2015} or second harmonic generation~\cite{Lee1967,Khurgin1995,Ruzicka2012,An2014}, which are observed both in bulk and low-dimensional structures.

In the present manuscript, we theoretically study the variation of the refractive index induced by dc electric current linear in current amplitude for the resonant optical transition between subbands in quantum well. The in-plane electric current in quantum well breaks inversion-symmetry and leads to nonequivalence of direction along and against the current.  As a result of this, the dielectric function of the quantum well can change depending on the orientation of the radiation wave vector with respect to the current direction. This phenomenon is the opposite to the well-known photon drag effect of electrons: the generation of a direct electric current caused by the absorption of radiation due to the transfer of photon momentum to the free charge carriers. Photon drag effect was  widely studied in both bulk semiconductors~\cite{Gibson1970,Danishevskii1970,ganichev2005book,Shalygin2017} and quantum wells~\cite{Luryi1987,grinberg1988,Ganichev2002,Shalygin2007,Stachel2014} 
 and used for characterization of semiconductor structures kinetic properties as well as ultrafast infrared detectors~\cite{Rogalski2010,ganichev2005book}.  
By analogy, the effect under study can be named the current drag of photons   (\effectname), since optical path length of radiation transmitted through quantum well changes in presence of dc current as if photons are ``dragged'' by electrons. 
A similar phenomenon caused by hole current at the optical transition between light-hole and heavy-hole subbands was previously discovered in bulk Ge~\cite{Vorobev2000}.

\section{Microscopic model}

\begin{figure}
\includegraphics[width=\linewidth]{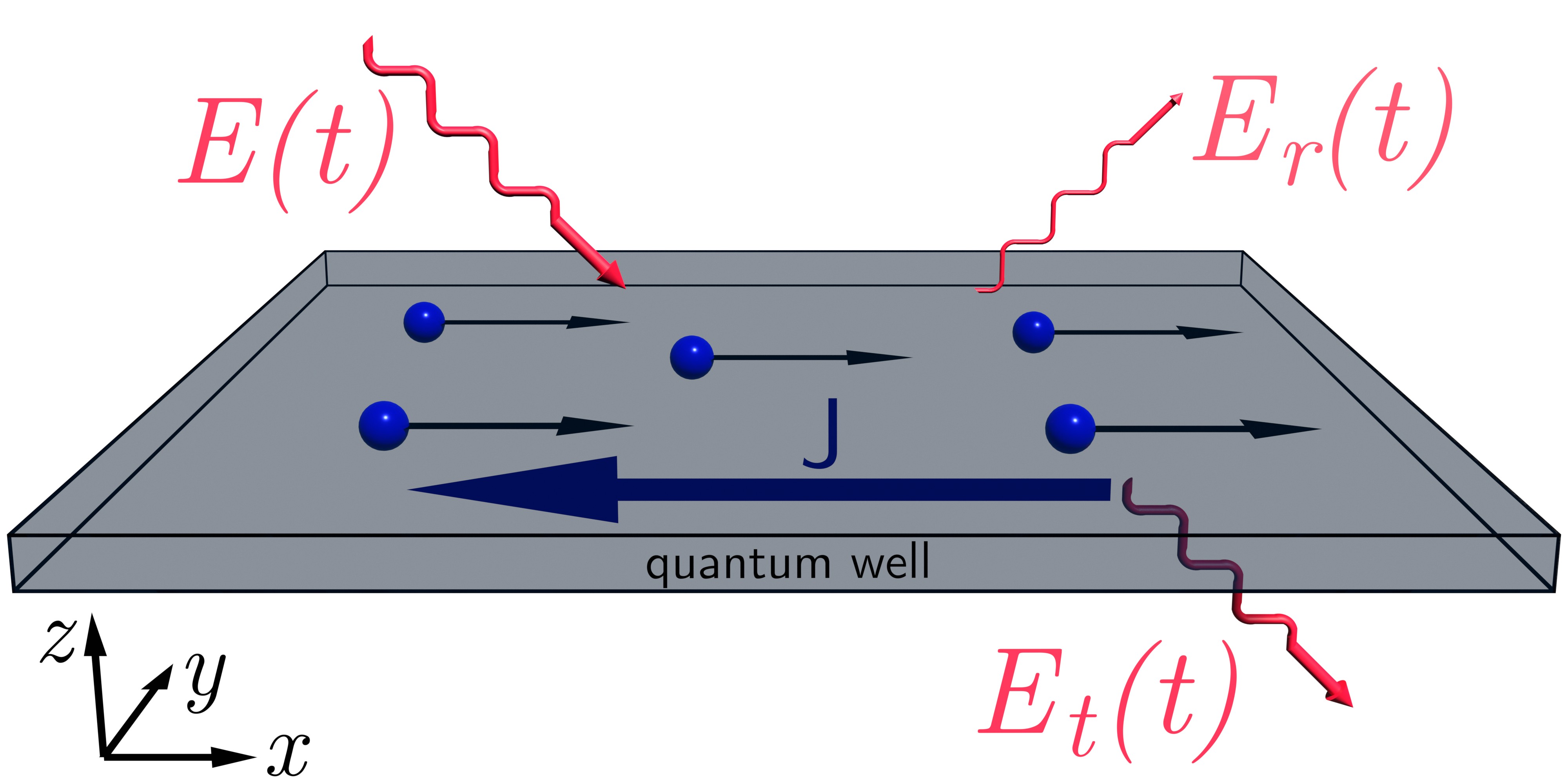}
\caption{
\label{geometry}
Figure schematically demonstrates the geometry of the problem. 
Incident radiation described by electric field $\bm{E}(t)$
is partially reflected and transmitted from quantum well with dc electric current, which are described by $\bm{E}_r(t)$ and $\bm{E}_t(t)$, respectively. 
}
\end{figure}

We start with the microscopic model to demonstrate physics behind the \effectname.
The geometry of the problem is shown in Fig.~\ref{geometry}. 
We consider direct optical transitions between the first and second subbands of the quantum well. 
For such transitions, the radiation frequency typically is in the infrared region.
Direct intersubband optical transitions are induced by the oblique and \textit{p}-polarized incident radiation, so its electric field has both in-plane and perpendicular to the quantum well components.
Dc electric current affects the radiation transition through the quantum well, which results in the dependence of the $E_t(t)$ and $E_r(t)$ on the current magnitude and direction for the fixed angle of incidence. To explain the reason for this dependence, we first turn to a formal description of the electronic states in the quantum well.
The Hamiltonian of electron in quantum well is given by
\begin{equation}
\hat{H}_0=\dfrac{\hat{\bm{p}}^2}{2 m^*}+U(z)\:,
\end{equation}
where $m^*$ is the electron effective mass, $U(z)$ is barrier potential. The solution of this well-known Shr\"odinger equation gives a series of electron subbands energies $E_i(k)=E_i^{(0)}+\hbar^2 k^2/(2 m^*)$
with corresponding wave functions $|i,\bm{k}\rangle$, in coordinate representation $$\langle \bm{r} | i,\bm{k}\rangle=\Psi_{i,\bm{k}}(\bm{r})=\varphi_i(z) \dfrac{e^{i \bm{k} \bm{r}_{\|}}}{\sqrt{S}}\:,$$
where $\bm{k}$ is in-plane electron wave vector 
and  $S$ is the surface  area of the quantum well.

\begin{figure}
\includegraphics[width=\linewidth]{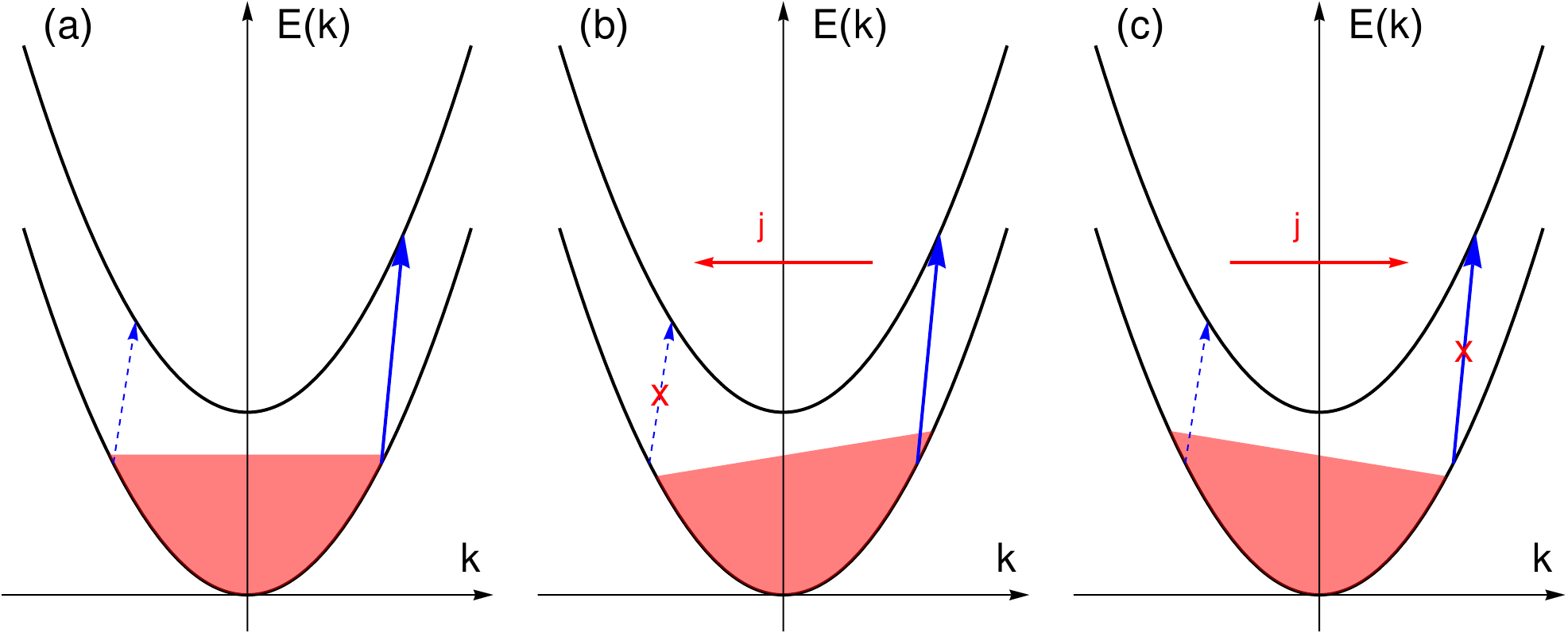}
\caption{
\label{model_picture}
Optical transition of the electrons in three cases:
(a) with zero net electric current,
(b) in the presence of current along $\bm{q}_{\|}$
and (c) with the opposite direction of current. Dashed and thick blue arrows demonstrate the contribution of transitions with different initial electron momentum to the dipole momentum of quantum well along $z$-axis.
}
\end{figure}

Radiation induces dipole moment along $z$-direction in quantum well oscillating with the frequency of the incident electromagnetic wave. Dipole moment density along $z$-direction is given by
\begin{equation}
\label{dipole_momentum}
\Pi_z=2\sum\limits_{\bm{k}} f_{\bm{k}} \Pi_{\bm{k}'\bm{k}}\:,
\end{equation}
where $f_{\bm{k}}$ is electron distribution function and $\Pi_{\bm{k}'\bm{k}}$ contribution from individual transitions of electrons with initial state in the first subband with wave vector $\bm{k}$  and final state in the second subband with wave vector $\bm{k}'$. For the direct optical transition the initial electron wave vector $\bm{k}$ differs from the final $\bm{k}'$  
by the in-plane component of radiation wave vector $\bm{q}_{\|}$ due to in-plane momentum conservation law.
 The contributions to the dipole moment induced by optical transitions is proportional to
\begin{equation}
\Pi_{\bm{k}'\bm{k}}\propto \dfrac{e z_{21}}{E_2(\bm{k}+\bm{q}_{\|})-E_1(\bm{k})-\hbar \omega-i \hbar \Gamma}\:,
\end{equation}
where $z_{21}$ is matrix element of coordinate between first and second subband, $e$ is electron charge,  $\hbar \omega$  is photon energy and $\Gamma$ is optical transition dephasing rate.  The difference between initial energy of electron and final is given by
\begin{equation}
\label{en_diff}
E_2(\bm{k}')-E_1(\bm{k})=E_{21}+\dfrac{\hbar^2}{2 m^*}
q_{\|}^2
+\dfrac{\hbar^2}{ m^*}
\bm{q}_{\|} \bm{k}\:,
\end{equation}
where $E_{21}=E_2^{(0)}-E_1^{(0)}$ is energy difference between the ground and second subbands, we assume that linewidth $\hbar \Gamma$ is much smaller than $E_{21}$.
Last term of the right-hand side of Eq.~\eqref{en_diff} shows that the contribution of the transition to the $\Pi_z$ depends on the relative direction of wave vectors $\bm{k}$ and $\bm{q}_{\|}$. 
If we set the direction of $\bm{q}_{\|}$ along $x$-axis the values of $\Pi_{\bm{k}'\bm{k}}$ will depend on the sign of $k_x$ which is schematically shown in Fig.~\ref{model_picture}(a) by thick and dashed blue arrows, thus there is a definite asymmetry of optical transitions.

The physics behind \effectname{}  can be most clearly demonstrated for the case of low temperatures for degenerated electron gas described by the Fermi step function.
Dc current leads to the redistribution of electrons in $k$-space, thus the current along $x$-direction turns off optical transition
with positive electron wave vectors close to the Fermi energy
due to the depopulating of states (see Fig.~\ref{model_picture}(b)), current with opposite direction, on the other hand, turns off transition with negative electron wave vectors (see Fig.~\ref{model_picture}(c)). As a result dipole momentum density given by Eq.~\eqref{dipole_momentum} changes with the change of the current direction, which leads to the fact that phase shift and amplitude of the electric field of the transmitted radiation depends on the relative direction of the radiation wave vector and electric current in quantum well.

\section{Theory}

Now we turn to the detailed theory of the effect.
The Hamiltonian of the electrons excited by electromagnetic wave is given by
\begin{equation}
\hat{H}=\hat{H}_0+\hat{V} e^{-i\omega t}+ \hat{V}^\dagger e^{i \omega t}\:,
\end{equation}
where $\hat{V}$ is perturbation induced by incident radiation. By solving equation for density matrix  one obtains for non-diagonal component
$
\rho_{21}(\bm{k})=\gamma_{21}(\bm{k})\exp(-i \omega t)\:,
$
where 
\begin{equation}
\label{gamma21}
\gamma_{21}(\bm{k})=\dfrac{-\langle 2 ,\bm{k}+\bm{q}_{\|}| \hat{V} | 1 ,\bm{k}\rangle}{\hbar \Delta \omega-i \hbar \Gamma} f^{(0)}(\bm{k})\:,
\end{equation}
$\hbar \Delta \omega=E_2(\bm{k}+\bm{q}_{\|})-E_1(\bm{k})-\hbar\omega$ is the offset frequency, we assume that without perturbation caused by radiation the second subband is completely empty and all electrons are in the first subband.
The interaction of electrons confined in quantum well with radiation
causes oscillations of both electronic charge  given by
\begin{equation}
\delta \rho_e(\bm{r})= e\varphi_1(z) \varphi_2(z) \overline{\gamma}e^{i \bm{q}_{\|}\bm{r}}e^{-i\omega t}+\text{c.c.} \:,
\end{equation}
and current along $z$-direction the main contribution to which has the form
\begin{equation}
j_z=\dfrac{-i \hbar e}{2 m^*}  \Phi(z) \overline{\gamma}e^{i \bm{q}_{\|}\bm{r}}e^{-i\omega t}+\text{c.c.} \:,
\end{equation}
here we also introduced terms $\Phi(z)=\varphi_1 (z) \partial_z \varphi_2 (z)-\varphi_2 (z) \partial_z \varphi_1 (z)$ and $\overline{\gamma}=2\sum_{\bm{k}} \gamma_{21}(\bm{k})/S$, factor $2$ stands for spin degeneracy. Without loss of generality, we set $\bm{q}_{\|}$ along $x$-axis, while dc current may have arbitrary direction. For \textit{p}-polarized  electromagnetic wave
nonzero components of electric and magnetic field are $E_x$, $E_z$ and $H_y$ and the fields themselves have the form
\begin{align*}
\bm{E}(\bm{r},t)&=[\bm{o}_x E_x(z)+\bm{o}_z E_z(z)]e^{i\bm{q}_{\|}\bm{r}-i \omega t}+\text{c.c.}\:,\\
\bm{H}(\bm{r},t)&=\bm{o}_y H_y(z)e^{i\bm{q}_{\|}\bm{r}-i \omega t}+\text{c.c.}\:.
\end{align*}
Maxwell equations for the components of electric induction, electric field and magnetic field
are given by
\begin{align}
\label{maxwell}
&i q_{\|} D_x(z)+\partial_z D_z(z)=4 \pi e\: \varphi_1(z) \varphi_2(z) \overline{\gamma}\:, \nonumber\\
&\partial_z E_x(z)-i q E_z(z)=\dfrac{i \omega}{c}H_y(z)\:,  \nonumber\\
&i q_{\|} H_y(z)=4\dfrac{\pi}{c}\dfrac{-i \hbar e}{2 m^*} \Phi(z) \overline{\gamma}+\dfrac{-i\omega}{c}D_z(z)\:,
\end{align}
electric induction and electric field are related by ${\bm{D}=\varepsilon_i \bm{E}}$, where $\varepsilon_i$ is permittivity of medium, $\varepsilon_1$ corresponds to the barrier  and  $\varepsilon_2$ is dielectric function of the material of the quantum well.
For simplicity we assume
that the quantum well has infinite barriers, such that the wave functions do not penetrate into the barriers.
In this case straightforward calculation of~\eqref{maxwell} yields equation for $E_z(z)$
\begin{equation}
\label{ez_equation}
\dfrac{2 \pi e \hbar}{m^* \omega \varepsilon_2} \overline{\gamma}
\left(\partial_z^2+\varepsilon_2\dfrac{\omega^2}{c^2} \right) \Phi(z)+
\left(\partial_z^2+k_{2,z}^2 \right)E_z(z)=0\:,\\
\end{equation}
where $k_{i,z}^2= \varepsilon_i\omega^2/c^2-  q_{\|} ^2$ is radiation wave vector along $z$-axis.
Electric field outside of the quantum well is described by 
\begin{align}
\label{inside_qw}
E_z(z)&=E_0 e^{i k_{1,z}z}+\mathcal{R} \:E_0   e^{-i k_{1,z}z}\:, &\quad z<0 \nonumber\\
E_z(z)&= \mathcal{T} \:E_0 e^{i k_{1,z}z}\:, \quad   &z>a
\end{align}
where $a$ is quantum well width, $\mathcal{R}$ and $\mathcal{T}$ are reflection and transmission coefficients, respectively. This coefficients allow to obtain the quantum well absorbance, which is given by $\eta=1-|\mathcal{T}|^2-|\mathcal{R}|^2$. 
Electric field inside quantum well for $0<z<a$ can be found using Green's function method
\begin{multline}
\label{outside_qw}
E_z(z)=A \:E_0  e^{i k_{2,z}z}+B \:E_0   e^{-i k_{2,z}z}   \\
   -\int dz'\:G(z-z') \dfrac{2 \pi e \hbar}{m^* \omega \varepsilon_2} \\
       \times\overline{\gamma}\left(\partial_{z'}^2+\varepsilon_2\dfrac{\omega^2}{c^2} \right) \Phi(z')\:, 
\end{multline}
where  $G(z-z')=e^{i k_{2,z}|z-z'|}/(2 i k_{2,z})$ is Green's function, coefficients $A$ and $B$ describe general solution and found from boundary conditions at $z=0$ and $z=a$, which are  $\varepsilon_i E_z(z)$ and $\partial_z E_z(z)$ being continuous.
By solving~\eqref{ez_equation} using~\eqref{inside_qw} and \eqref{outside_qw}
one obtains, that reflection and transmission are given by
\begin{equation}
\label{t_intermediate}
\mathcal{T}=1+\mathcal{R}=1+\dfrac{i}{k_1}\dfrac{\pi e \hbar}{m^* \omega \varepsilon_2} \dfrac{\overline{\gamma}}{E_0}  q_{\|} ^2 \int_0^a dz\: e^{i k_2 z} \Phi(z)\:,
\end{equation}
here the presence of electric current in the quantum well is taken into account in $\overline{\gamma}$. Latter is found by summing over all $\bm{k}$ of~\eqref{gamma21}, where the part of the Hamiltonian describing perturbation is given by
\begin{equation}
\label{interaction}
\hat{V}=-\dfrac{e}{c}\dfrac{\hat{p_z}A_z(z)+A_z(z)\hat{p_z}}{2m^*}+\dfrac{\partial U_{xc}[n(\bm{r})]}{\partial n(\bm{r})} \delta n(\bm{r})\:,
\end{equation}
where first term of the right-hand side stands for interaction of electrons with alternating electric field and second term
takes into account exchange-correlation effects.
Here we chose a gauge $A_z(z)=c E_z(z)/(i \omega)$, where electric potential is static and interaction of electrons is described by the means of vector potential.  Equation~\eqref{interaction} takes into account depolarization effect  (see Refs.~\cite{ando1982,ivchenko2005book}), since $E_z(z)$ consists not only of the field of incident radiation, but also of the field induced by oscillations of electron density along $z$-direction. $U_{xc}[n(\bm{r})]$ in the Eq.~\eqref{interaction} is the exchange-correlation energy in the local density approximation  (for details see  Refs.~\cite{ando1982,Bloss1989,Schneider2006}), $n(\bm{r})$ is electron gas density and $\delta n(\bm{r})=\delta \rho_e(\bm{r})/e$ is oscillating part of electron density. Self consistent equation for $\gamma$ has the form
\begin{multline}
\overline{\gamma}= \sum\limits_{\bm{k}} 
\dfrac{2}{S} \dfrac{f^{(0)}(\bm{k})}{\hbar \Delta \omega- i \hbar \Gamma}
\\\times\biggl[
\dfrac{\hbar e}{2 m^* \omega}
\biggl(
\dfrac{\varepsilon_1}{\varepsilon_2} E_0 \int\limits_0^a dz \:\Phi(z)
+ \dfrac{2 \pi e \hbar}{m^* \omega \varepsilon_2} \overline{\gamma} \int\limits_0^a dz \:\Phi^2(z)
\biggr)
\\-\overline{\gamma} \int\limits_0^a dz
\varphi_1^2(z)\varphi_2^2(z) \dfrac{\partial U_{\text xc}[n(z)]}{\partial n}
\biggr]\:.
\end{multline}
To describe electron distribution function in the presence of dc electric current we use drift velocity ansatz
\begin{equation}
f(\bm{k})=f^{(0)}(\bm{k}-\bm{k}_{\text{dr}})
=\dfrac{1}{e^{\dfrac{\dfrac{\hbar^2 (\bm{k}-\bm{k}_{\text{dr}})^2}{2 m^*}-E_F}{T}}+1}\:,
\end{equation}
where $E_F$ is the Fermi energy, $T$ is temperature in energy units, $N_e$ total electron gas density, $\hbar \bm{k}_{\text{dr}}=m^*  \bm{v}_{\text{dr}}$ and $\bm{v}_{\text{dr}}$
is electron gas drift velocity. The relation between drift velocity and electric current is given by $j=e N_e v_{\text{dr}}$.
Leaving leading correction terms in denominator, which are linear both in $\bm{q}_{\|}$ and $\bm{v}_{\text{dr}}$
we obtain
\begin{equation}
\label{gamma_final}
\overline{\gamma}= 
\dfrac{
\dfrac{\hbar e}{2 m^* \omega} E_0 \dfrac{\varepsilon_1}{\varepsilon_2} N_e \int\limits_0^a dz \:\Phi(z)
}
{
\tilde{E}_{21}+\hbar \bm{q}_{\|} \cdot \bm{v}_{\text{dr}}-\hbar  \omega- i \hbar \Gamma
}\:,
\end{equation}
where 
$\tilde{E}_{21}^2=E_{21}^2(1+\alpha-\beta)$ is shifted resonance energy,
\begin{equation}
\alpha=
 - \dfrac{ 8\pi N_e e^2 }{ \varepsilon_2 E_{21}}  \int\limits_0^a dz \left(\int\limits_0^z dz' \varphi_1(z') \varphi_2(z')\right)^2
\end{equation} describes depolarization effect~\cite{ivchenko2005book}
and
\begin{equation}
\beta=-\dfrac{2 N_e}{E_{21}}\int\limits_0^a dz \:\varphi_1^2(z)\varphi_2^2(z) \dfrac{\partial U_{\text xc}(n)}{\partial n}
\end{equation}
exchange-correlation effect~\cite{Bloss1989}.
Finally substitution of~\eqref{gamma_final} into Eq.~\eqref{t_intermediate}
leads 
to the transmission coefficient given by
\begin{equation}
\label{transmission_final}
\mathcal{T}=1+i \zeta\dfrac{e^2 \hbar}{2 m^* c}    \dfrac{\varepsilon_1^2}{\varepsilon_2^2 \sqrt{\varepsilon_1}} \dfrac{\sin^2 \theta}{\cos \theta} 
\dfrac{1}{\pi}
N_e
\dfrac{
 1
}
{
\tilde{E}_{21}+\hbar  \bm{q}_{\|}  \cdot \bm{v}_{\text{dr}}-
\hbar  \omega- i \hbar \Gamma
}\:,
\end{equation}
and absorbance given by
\begin{equation}
\label{absorption_final}
\eta= \zeta\dfrac{e^2 \hbar}{m^* c} N_e   \dfrac{\varepsilon_1^2}{\varepsilon_2^2 \sqrt{\varepsilon_1}} \dfrac{\sin^2 \theta}{\cos \theta} 
\dfrac{1}{\pi}
\dfrac{
\hbar \Gamma
}
{
(\tilde{E}_{21}+\hbar  \bm{q}_{\|}  \cdot \bm{v}_{\text{dr}}-
\hbar  \omega)^2+ (\hbar \Gamma)^2
}\:,
\end{equation}
where \[
\zeta=
 \left[\dfrac{\pi^2  \hbar^2}{m^{*} \hbar \omega} \left(\int\limits_0^a dz \:\Phi(z)\right)^2 \right]
\] is numerical coefficient. For the constant barrier potential inside the quantum well (particle in a box) close to the intersubband resonance frequency ${\zeta=512/27}$.
It is worth noting that
finite height of the barriers, static 
Hartree potential and exchange-correlation effects alter
subband energy difference $E_{21}$,
 electron wave function and coefficient $\zeta$. This effects are widely discussed in the literature~\cite{Bandara1988,Bloss1989,Schneider2006}, but
not related to the \effectname{} and not in the focus of  the present manuscript.

\section{Disscussion}

Considering optical properties quantum well can be treated as a uniaxial crystal
where in–plane permittivity tensor  components differ from perpendicular to the quantum well $\varepsilon_{zz}$.
The \effectname{} manifests itself as a shift in resonance frequency as can be seen from Eqs.~\eqref{transmission_final} and~\eqref{absorption_final}.
As a result the dielectric constant $\varepsilon_{zz}$ and refractive index of quantum well 
changes if an electric current is present in the quantum well.
For \effectname{} the permittivity tensor component along $z$-direction in the slab model~\cite{ivchenko2005book,steed2013} is given by
\begin{equation}
\label{permitivity}
\varepsilon_{zz}=\varepsilon_2+
 \zeta \dfrac{e^2 N_e}{m^* a_{eff}} \dfrac{1}{\omega}
\dfrac{1}{\pi}\dfrac{1}{\Delta\tilde{\omega}+ \bm{q}_{\|} \cdot \bm{v}_{\text{dr}}-i \Gamma}
\:,
\end{equation}
 where ${ \Delta\tilde{\omega}=\tilde{E}_{21}/\hbar-  \omega}$ and $a_{eff}\sim a$ is the effective thickness of the transition layer.

Since permittivity tensor depends on $\bm{v}_{\text{dr}}$, 
by altering dc current magnitude and direction one can change the optical path of transmitted radiation through quantum well.
The difference between the phase shift
in the presence of dc current and when current is absent is a measurable quantity that can be observed experimentally.
This addition phase shift caused by \effectname{} is designated $\delta \varphi$ and has the form
\begin{multline}
\label{effect_phase_shift}
\delta \varphi=-\zeta\dfrac{e^2 \hbar}{2 m^* c}  N_e  \dfrac{\varepsilon_1^2}{\varepsilon_2^2 \sqrt{\varepsilon_1}} \dfrac{\sin^2 \theta}{\cos \theta} 
\hbar  \bm{q}_{\|} \cdot \bm{v}_{\text{dr}} \\\times
\dfrac{1}{\pi}
\dfrac{
[ (\tilde{E}_{21}-
\hbar  \omega)^2-(\hbar \Gamma)^2]
}
{
[(\tilde{E}_{21}-
\hbar  \omega)^2+ (\hbar \Gamma)^2]^2
}\:,
\end{multline}
here we assumed total phase shift $\varphi\approx\Im(\mathcal{T}-1)$, which is typically relevant 
for quantum wells since absorption of single quantum well is weak.
\begin{figure}
\includegraphics[width=\linewidth]{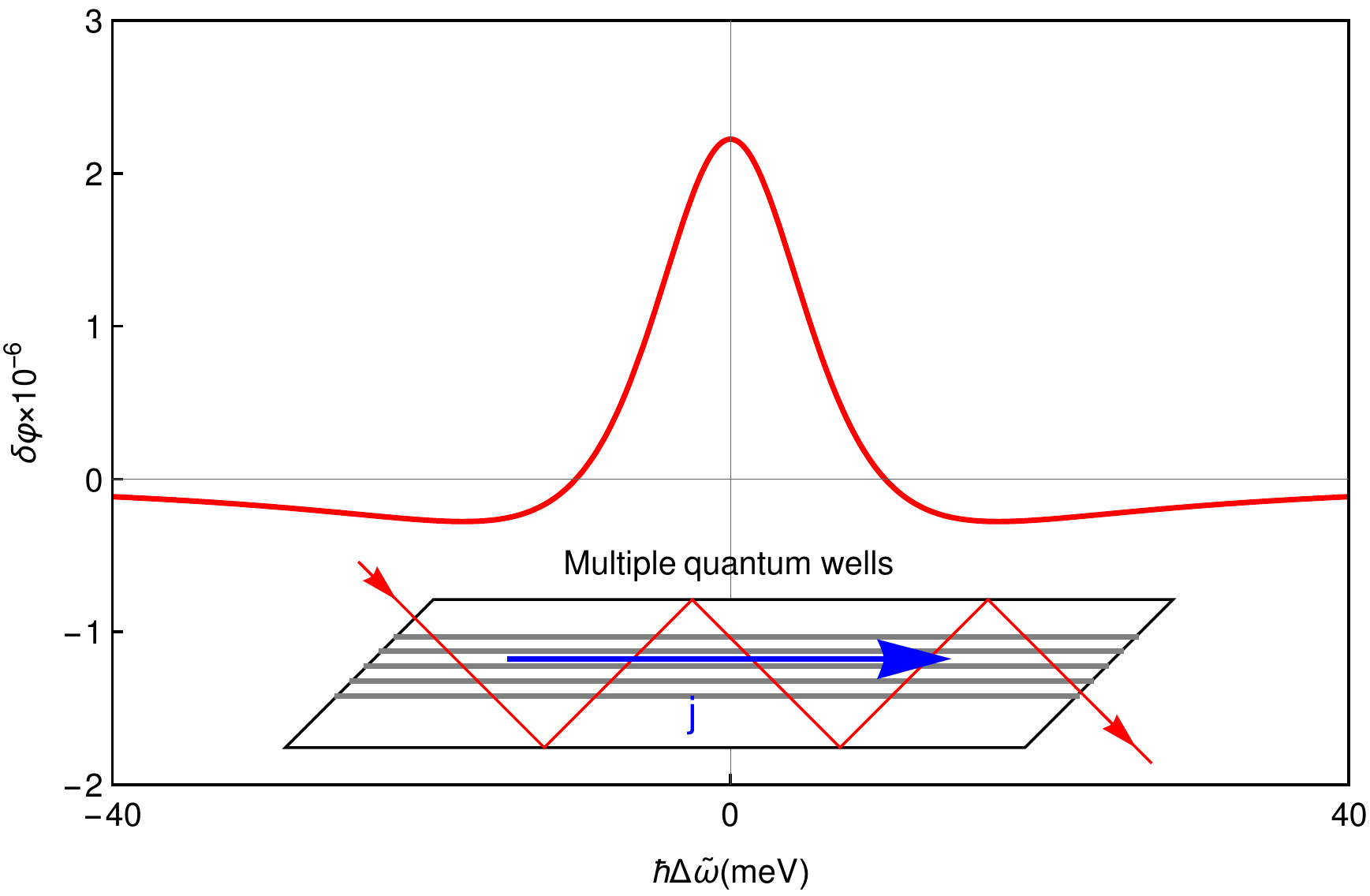}
\caption{
\label{figure_phase_change}
Figure shows dependence of additional phase shift induced by \effectname{} on the radiation frequency for GaAs/AlAs quantum well.
Curve is calculated after Eq.~\eqref{effect_phase_shift} for quantum well width $a=10$~nm, 
dephasing rate $\hbar \Gamma=10$~meV, electron density
$N_e=5\cdot 10^{11}$~cm$^{-2}$ and
drift electron velocity
${v_{\text{dr}}=10^6}$~cm/s.
The inset shows the setup of multiple-reflection wave-guide geometry, which can be used for detecting of the \effectname{}.
}
\end{figure}
The dependence of $\delta \varphi$ on radiation frequency is shown in Fig.~\ref{figure_phase_change}. Phase shift induced by \effectname{} exhibits maximum exactly  at intersubband resonance position $\tilde{E}_{21}=\hbar  \omega$, it changes sign at $\hbar \Delta\tilde{\omega}=\pm\hbar \Gamma$ and decays to zero away from resonance.

To increase the phase shift $\delta \varphi$  the sample 
 with multiple quantum wells in  wave-guide geometry can be used (see the inset in Fig.~\ref{figure_phase_change}).
For $N$ quantum wells and $M$ reflections phase shift caused by 
\effectname{} is $N \cdot M \cdot \delta \varphi$. Modern technology allows to grow structure with both ${N\propto 100}$ and ${M\propto 100}$, which yields the induced by \effectname{} phase shift $\sim 1^\circ$, which can be reasonably well observed in the experiment.

\begin{figure}
\includegraphics[width=\linewidth]{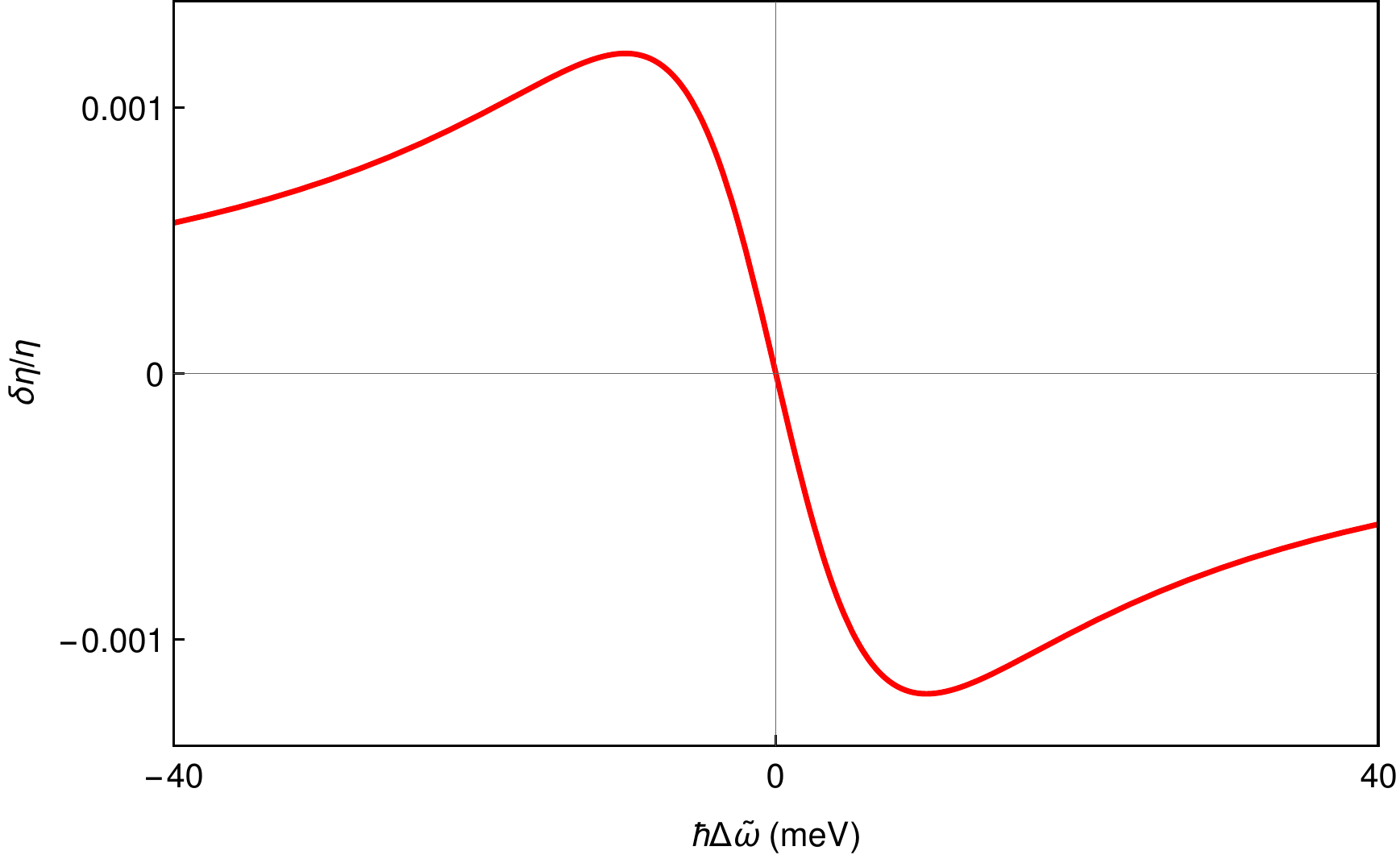}
\caption{
\label{figure_absorption}
The dependence of relative absorbance change caused by \effectname{} on the radiation frequency for GaAs/AlAs quantum well.
Curve is calculated after Eq.~\eqref{effect_absorption} for quantum well width $a=10$~nm, 
dephasing rate $\hbar \Gamma=10$~meV, electron density
$N_e=5\cdot 10^{11}$~cm$^{-2}$ and
drift electron velocity
${v_{\text{dr}}=10^6}$~cm/s.
}
\end{figure}

Another quantity which can be experimentally observed is absorbance. Relative correction to the absorbance induced by \effectname{} is given by
\begin{equation}
\label{effect_absorption}
\dfrac{\delta \eta}{\eta}=-2
\dfrac{\hbar  \bm{q}_{\|}  \cdot \bm{v}_{\text{dr}}\left(\tilde{E}_{21}-
\hbar  \omega\right)}{(\tilde{E}_{21}-
\hbar  \omega)^2+ (\hbar \Gamma)^2}\:.
\end{equation}
The dependence of $\delta \eta/\eta$ is shown in Fig.~\ref{figure_absorption}, the main features of relative correction to the absorbance
are as well observed at $\hbar \Delta \tilde{\omega}=0$  and $\hbar \Delta \tilde{\omega}=\pm \hbar \Gamma$.
However, unlike the phase shift caused by \effectname{} $\delta\eta/\eta$
is zero at intersubband resonance and has maximum at detuning $\hbar \Delta \tilde{\omega}=\pm \hbar \Gamma$.
The characteristic value of $\delta \eta/\eta$ of $0.1$\% is within reach of modern experimental capabilities.

Finally we point out that shift of resonance position $\hbar  \bm{q}_{\|}  \cdot\bm{v}_{\text{dr}}$ is actually exactly matches the Doppler shift in the frame of reference moving with $\bm{v}_{\text{dr}}$, which further supports the title of ``current drag of photons''.

\section{Summary} 
\label{summary}

Microscopic theory of the drag of photons by electric current in quantum wells have been developed. It has been shown that absorbance, transmission, and reflection coefficients are affected by dc electric current in quantum well. 
We demonstrated that  the \effectname{} is due to the asymmetry of intersubband transitions, this asymmetry appears by taking into account the shift of electron distribution in momentum space with respect to the wave vector of light.
 The main feature of the \effectname{} is  the shift of intersubband resonance position by the value of ${\hbar \bm{q}_{\|} \cdot \bm{v}_{\text{dr}}}$.
 The corrections caused by \effectname{} to the phase shift and radiation absorption are estimated for GaAs based quantum wells.

\section{Acknowledgments}
\label{acknow}

This work has been supported by the RFBR, project number 19-02-00273. G.V.B. also acknowledges support by the Foundation for the Advancement of
Theoretical Physics and Mathematics ``BASIS''.

\bibliography{ref4}
\end{document}